\documentclass[sigconf]{acmart}

\usepackage[T1]{fontenc}

\usepackage{amsmath}

\usepackage{amssymb}

\usepackage{bm}

\usepackage{graphicx}
\usepackage{booktabs}
\usepackage{multirow}

\usepackage{enumitem}

\usepackage{xcolor}

\usepackage[misc]{ifsym}

\usepackage[linesnumbered,ruled,vlined]{algorithm2e}

\usepackage{subcaption}
\usepackage{soul}
\usepackage{bbding}
\AtBeginDocument{%
  }

\setcopyright{acmlicensed}
\copyrightyear{2018}
\acmYear{2018}
\acmDOI{XXXXXXX.XXXXXXX}
\acmConference[Conference acronym 'XX]{Make sure to enter the correct
  conference title from your rights confirmation email}{June 03--05,
  2018}{Woodstock, NY}
\acmISBN{978-1-4503-XXXX-X/2018/06}

\begin{document}

\title{Behavior-Aware Dual-Channel Preference Learning for Heterogeneous Sequential Recommendation}

% \author{Anonymous Author(s)}

\author{Jing Xiao}
\affiliation{%
  \institution{College of Computer Science and Software Engineering, Shenzhen University}
  \city{Shenzhen}
  \country{China}}
\email{xiaojing2022@email.szu.edu.cn}

\author{Dongqi Wu}
\affiliation{%
  \institution{College of Computer Science and Software Engineering, Shenzhen University}
  \city{Shenzhen}
  \country{China}}
\email{wudongqi2023@email.szu.edu.cn}

\author{Liwei Pan}
\affiliation{%
  \institution{College of Computer Science and Software Engineering, Shenzhen University}
  \city{Shenzhen}
  \country{China}}
\email{panliwei2023@email.szu.edu.cn}

\author{Yawen Luo}
\affiliation{%
  \institution{College of Computer Science and Software Engineering, Shenzhen University}
  \city{Shenzhen}
  \country{China}}
\email{2410104006@mails.szu.edu.cn}

\author{Weike Pan}
\affiliation{%
  \institution{College of Computer Science and Software Engineering, Shenzhen University}
  \city{Shenzhen}
  \country{China}}
\email{panweike@szu.edu.cn}

\author{Zhong Ming}
\affiliation{%
  \institution{College of Computer Science and Software Engineering, Shenzhen University}
  \city{Shenzhen}
  \country{China}}
\email{mingz@szu.edu.cn}

% \author{}
% \affiliation{%
%   \institution{}
%   \city{}
%   \country{}}
% \email{}


\begin{abstract}
Heterogeneous sequential recommendation (HSR) aims to learn dynamic behavior dependencies from the diverse behaviors of user-item interactions to facilitate precise sequential recommendation. Despite many efforts yielding promising achievements, there are still challenges in modeling heterogeneous behavior data. One significant issue is the inherent sparsity of a real-world data, which can weaken the recommendation performance. Although auxiliary behaviors (e.g., clicks) partially address this problem, they inevitably introduce some noise, and the sparsity of the target behavior (e.g., purchases) remains unresolved. Additionally, contrastive learning-based augmentation in existing methods often focuses on a single behavior type, overlooking fine-grained user preferences and losing valuable information. To address these challenges, we have meticulously designed a behavior-aware dual-channel preference learning framework (BDPL). This framework begins with the construction of customized behavior-aware subgraphs to capture personalized behavior transition relationships, followed by a novel cascade-structured graph neural network to aggregate node context information. We then model and enhance user representations through a preference-level contrastive learning paradigm, considering both long-term and short-term preferences. Finally, we fuse the overall preference information using an adaptive gating mechanism to predict the next item the user will interact with under the target behavior. Extensive experiments on three real-world datasets demonstrate the superiority of our BDPL over the state-of-the-art models. 
% The datasets, scripts and source codes of the baselines and our BDPL are available at https://anonymous.4open.science/r/BDPL-D481/.

\end{abstract}





\maketitle

\section{Introduction}
Sequential recommendation (SR), which exploits the sequential nature of user-item interactions to strive for more accurate and personalized recommendations, has emerged as a prominent and increasingly focused research area in recommender systems\cite{FPMC, Caser, SASRec}. 
While traditional sequential recommendation primarily focuses on single behavior types, such as clicks or purchases, real-world user interactions are often more complex, involving multiple behaviors (e.g., views, carts, purchases) across different items. To capture a more comprehensive understanding of user preferences, great efforts have been made on heterogeneous sequential recommendation (HSR), which aims to model diverse behavior interactions to mine diverse sequential patterns and interrelated behavior dependencies.

Up to now, research on HSR has made significant progress\cite{add_3,add_4}. 
From a technical perspective, these approaches can be broadly classified into three categories, including recurrent neural network (RNN)-based, Transformer-based, and graph neural network (GNN)-based methods. 
Among them, RNN-based methods \cite{RIB, BINN,DyMus} introduce heterogeneous information by integrating behavior embeddings at the input layer. Transformer-based \cite{BAR,MBSTR,PBAT}  methods differentiate the impact of various behaviors on user preferences through fine-grained modeling of specific behavior subsequences. GNN-based methods \cite{mbht,MGNN-SPred,DSUIL} capture cross-behavior transitions by constructing global or local graphs. 
Though effective, these methods often struggle to maintain their original recommendation accuracy when dealing with sparse datasets. In other words, sparse interactions hinder the model’s ability to capture the behavior multiplicity and dependencies, thereby impairing its performance.

Motivated by the achievements of contrastive learning, researchers have sought to apply it to recommender systems to address the data sparsity issue through data augmentation\cite{MBASR,BLADE,DiffAMB}. 
The core idea of contrastive learning is to devise semantically similar and dissimilar self-supervised signals to enhance representation learning.
By pulling the positive pairs closer together and pushing the negative pairs further apart, the model can uncover potential correlations behind rich structure in heterogeneous data, thus deepening the understanding of user preferences.
For instance, $S^3$-Rec \cite{s3rec} designs four self-supervised learning objectives based on a self-attentive structure and uses maximum mutual information (MIM) to enhance data representation for sequential recommendation. CL4SRec \cite{CL4SRec} and CoSeRec \cite{CoSeRec} construct new samples under multiple augmentation operations from the raw data and apply contrastive learning to improve the backbone model performance. Subsequently, DuoRec \cite{DuoRec} treats sequences with the same target behavior as positive pairs, while HPM \cite{HPM} employs a dual-channel Transformer with two contrastive learning mechanisms to model user preferences at different levels.
Despite the promising results of these contrastive learning paradigms in refining user preferences, we argue that they still show two common limitations.

\noindent\textbf{Heterogeneous Information under Multiple Behaviors.} These methods fail to consider the heterogeneity of user behaviors during modeling. Different types of behaviors often carry distinct semantic and preference information. Treating these heterogeneous behaviors uniformly may obscure their differences, preventing the model from accurately capturing the complex user intentions.

\noindent\textbf{Semantic-rich Self-Supervised Signal Construction.} Most of these models rely on explicitly generating new samples or modifying the model structure to obtain new representations. Though these methods can somewhat enrich the training data, they often fail to construct self-supervised signals from the semantic level of user preferences in a deeper way. This inevitably introduces noise and increases data inconsistency, which could lead to sub-optimal performance. 

% 完全依赖新构建
To tackle the aforementioned drawbacks, we propose a novel dual-channel contrastive preference learning model for HSR.
It aims to anchor users' long-term and short-term preferences and design a contrast enhancement mechanism for two types of preferences. 
Specifically, our BDPL distinguishes itself through the following three key stages. \textbf{Firstly, to address the first limitation}, for different behaviors, we introduce tailored behavior transition patterns, which include both homogeneous and heterogeneous relations, to construct multiple behavior-aware subgraphs. Notably, when capturing heterogeneous behavior transitions, we only consider transitions from auxiliary behaviors to the target behavior. This approach helps the model focus more on predicting the target behavior, aligning with the core problem of our study. Moreover, for each target behavior, we associate it with a sequence of preceding auxiliary behaviors, as we posit that auxiliary behaviors occurring between two target behaviors may directly relate to the next target behavior, rather than merely considering the auxiliary behavior adjacent to the target behavior.
Following this, we design a behavior-aware graph encoder to aggregate item representations under different behaviors. The encoder features a cascaded network from the target behavior to auxiliary behaviors, aiming to purify and guide the learning of auxiliary behaviors based on the representation of the target behavior. Overall, we hope these behavior-aware subgraphs can effectively distill the rich transition relationships between behaviors, making them more suitable for target behavior-oriented (i.e., purchase) recommendation. \textbf{Secondly, to address the second limitation}, we focus on modeling user preferences through a dual-path approach, employing different network structures to capture long-term and short-term preferences. For each type of preferences, we explicitly propose another view of representation. By bringing the representations of the same preferences closer under these two views, we enhance the modeling of user preferences, implementing a preference-level contrastive learning enhancement. \textbf{Finally}, we use an adaptive gating mechanism to fuse long-term and short-term preferences, resulting in the final preference representation for prediction.

We summarize the main technical contributions and novelties as follows:
\vspace{-0.3em}
\begin{itemize}[leftmargin=*]
    \item We customize behavior-aware multi-hop subgraphs to explore diverse sequential patterns under different behaviors, where each subgraph incorporates both homogeneous and heterogeneous behavior transition relationships, with a focus on transitions oriented towards the target behavior.
    \item We propose a novel cascaded graph convolutional structure that flows from the target behavior to auxiliary behaviors, aiming to leverage the target behavior to facilitate information propagation for the auxiliary behaviors.
    \item We encode users' long-term and short-term preferences from different perspectives. Specifically, we use an adaptive preference-aware fusion module to integrate these preferences, and introduce a dual-channel contrastive learning mechanism to enhance preference learning.
    \item We conduct extensive experiments on three real-world datasets. Empirical results show that our BDPL can outperform the state-of-the-art baselines, and further analysis confirms its effectiveness and superiority.
\end{itemize}

\section{Related Work}
\subsection{Sequential Recommendation}
\subsubsection{Homogeneous Sequential Recommendation}
A key advancement in sequential recommendation is homogeneous sequential recommendation, which captures dynamic user preferences by modeling their historical behavior sequences for next-item prediction. 
Early studies use Markov chains (MCs) to learn item-item transitions \cite{FPMC}.  
With the rapid development of deep neural networks, a series of deep learning-based models have emerged, achieving substantial success. 
Among them, the RNN-based model GRU4Rec \cite{GRU4Rec} utilizes GRU to encode users' dynamic preferences. 
Based on this, several improved RNN-based models have been developed \cite{GRU4Rec+, GRU4Rec-DWell}. 
Simultaneously, convolutional neural networks (CNNs) have also shown great potential in modeling sequential data \cite{Caser, NextItNet}. For example, 
the CNN-based model Caser \cite{Caser} 
treats consecutive items in interaction sequences as images and uses both vertical and horizontal convolutional filters to exploit item-item correlations. 
Moreover, attention-based models enhance the ability to capture long-term dependencies \cite{SASRec, DDualSE}, while GNN-based models use elaborate graph structures to capture complex transitions between neighboring items \cite{gnn1,gnn25}. 
However, most of these models are tailored for a single type of user interaction and fail to consider the diversity of user behaviors, which can provide valuable insights into user intentions.

\subsubsection{Heterogeneous Sequential Recommendation}
Heterogeneous sequential recommendation (HSR) leverages multiple user-item interaction types to track interest drift\cite{M-GPT,add_1}. Recent studies on HSR largely rely on deep learning, with RNN being the first to be applied to.
RIB \cite{RIB} introduces behavior-specific embeddings to encode information for different behaviors, and utilizes GRU to capture sequential information.
BAR \cite{BAR} proposes a generic behavior-aware attention framework that enables SBSR models to address MBSR scenarios effectively. 
NextIP \cite{NextIP} performs a purchase prediction task by modeling the transitions between the auxiliary and target behaviors.
 MBSTR \cite{MBSTR} and PART \cite{PBAT} both employ Transformers to model multi-behavior interaction sequences, thus capturing dependencies between items and behavior heterogeneity. 
 Recently, GNNs have been used to capture potential user intent in graph-structured data for recommendation tasks via cross-layer message information propagation, bringing large performance gains. 
 For example,  
 GPG4HSR \cite{GPG4HSR} employs a global graph to capture transitions of heterogeneous behaviors and designs a local graph to incorporate personalized user context information. 
 GHTID \cite{GHTID} further adopts a behavior-aware attention mechanism module to capture specific behavioral patterns in the item-to-item transition graph.
 In addition, BMLP \cite{BMLP} fuses item and behavior embeddings as inputs and uses only an MLP to capture diverse behavioral features and sequential patterns.
Although these methods have made promising progress, their performance on sparse datasets remains unsatisfactory, as data sparsity often suppresses their performance.

\subsection{Contrastive Sequential Recommendation}
Inspired by the advances in computer vision and natural language processing \cite{cvcl2, cvcl3}, contrastive learning (CL) has been widely introduced to enhance representation learning in SR via data augmentation, which is regarded as an effective technique for addressing the issue of data sparsity. 
$S^3$-Rec \cite{s3rec} adopts mutual information to maximize attributes, items, segments and sequences.
CL4SRec \cite{CL4SRec} employs sequence-level augmentation to construct different positive pairs for a same sequence.
CoSeRec \cite{CoSeRec} builds upon CL4SRec by introducing data-level 
augmentations.
DuoRec \cite{DuoRec} designs model-level augmentation based on Dropout. It also proposes a novel concept that sequences with the same target item should be treated as hard positive samples, allowing for the construction of semantically rich sample pairs.
DCRec \cite{DCRec} constructs an item-item transition graph and a user-item co-interaction graph to model user preferences from sequential and collaborative perspectives, respectively, and enhances these representations with adaptive cross-view contrastive learning.
Our work is in line with this class of methods. Specifically, we design an explicit preference learning framework that models long-term and short-term user preferences from two distinct perspectives, and enhance sequence representation through dual-channel contrastive learning.

\section{PRELIMINARIES}
In this section, we first give the formal definition of heterogeneous sequential recommendation (HSR). Then, we give an overview of our proposed BDPL. 

\subsection{Problem Definition}
Let $\mathcal{U}=\{u_1, u_2, \cdots, u_{|\mathcal{U}|}\}$ and $\mathcal{V}=\{v_1, v_2, \cdots, v_{|\mathcal{V}|}\}$ denote the sets of users and items, respectively. Suppose there are ${|\mathcal{B}|}$ types of behaviors, denoted by $\mathcal{B}=\{b_1, b_2, \cdots, b_{|\mathcal{B}|}\}$. For each user $u \in \mathcal{U}$, his or her behavior-aware interaction sequence in chronological order is defined as $\mathcal{S}_u=\{(v_1, b_1), \cdots, (v_t, b_t), \cdots, (v_{|\mathcal{S}_u|}, b_{|\mathcal{S}_u|})\}$. Each (item, behavior) pair in $\mathcal{S}_u$ is composed of the $t$-th interacted item $v_t \in \mathcal{V}$ and the corresponding behavior type $b_t \in \mathcal{B}$. Following previous studies \cite{GPG4HSR}, we take purchase $p$ as $\textit{target behavior}$, while considering other behaviors (i.e., view, add-to-favorite, and add-to-cart) as \textit{auxiliary behaviors}, denoted as examination $e$.
Given a user's heterogeneous interaction sequence, our goal is to predict the next item $v_{|\mathcal{S}_u|+1}\in \mathcal{V}$ that user $u$ is most likely to purchase, which can be formulated as follows:
    \begin{equation}
    \underset{v_i \in \mathcal{V}}{\arg \max } P\left(v_{\left|\mathcal{S}_u\right|+1}=v_i \mid \mathcal{S}_u\right)
    \end{equation}

% +++++++++++++++++Graph Construction+++++++++++++++++++++++

\subsection{Overview of Our Solution}
\begin{figure*}[!htbp] 
    \centering 
    \setlength{\abovecaptionskip}{0.cm}
    \includegraphics[width=1\textwidth]{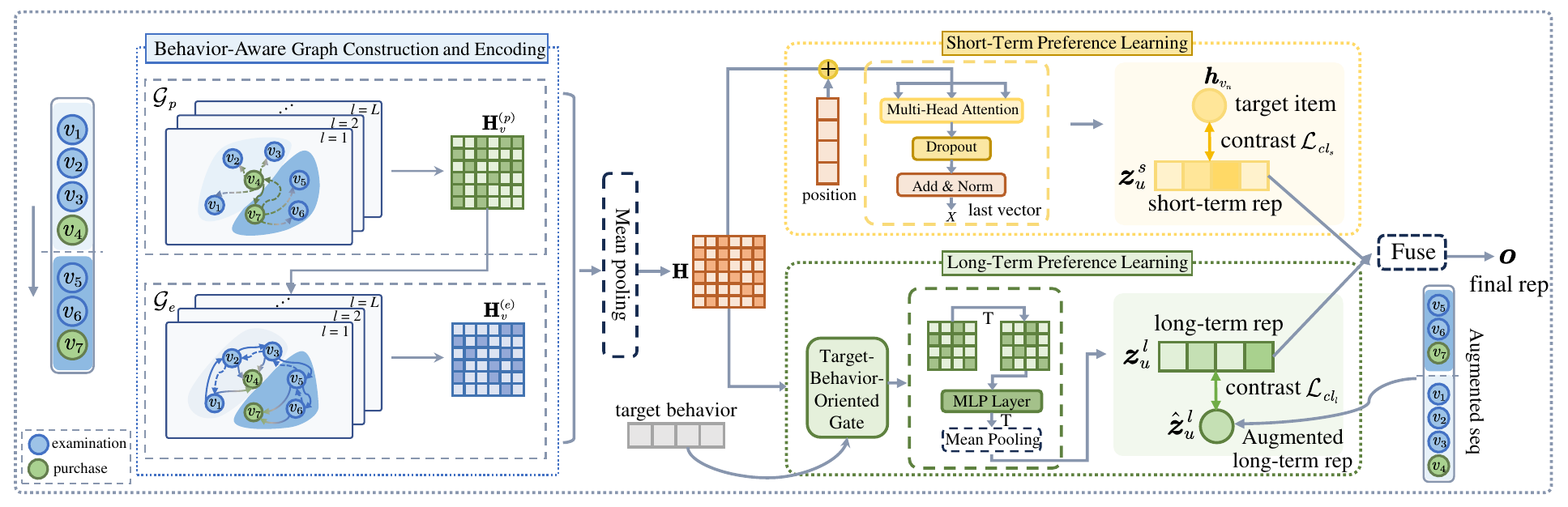} 
    \caption{The overall architecture of the proposed BDPL model. The blue circles represent auxiliary behaviors (examination), the green circles represent target behaviors (purchase), a solid arrow from $v_i$ to $v_j$ and a dashed arrow from $v_j$ to $v_i$ both denote a transition from $v_i$ to $v_j$ in the original sequence, and ``T'' denotes transposition.} 
    \label{framework} %用于文内引用的标签
\end{figure*}
% \vspace{-1cm}
The overall framework of our BDPL is shown in Figure \ref{framework}. In general, we design a dual-channel contrastive learning mechanism to enhance the modeling of both long-term and short-term user preferences. For each preference, we generate the representation from two perspectives and employ contrastive learning to minimize the distance between them.
Specifically, we first construct behavior-aware subgraphs, which include both homogeneous and heterogeneous transition relations, and use a behavior-aware graph encoder to capture rich behavior semantic contexts through multiple iterations. Then, we employ a behavior-integrated attention encoder to learn the user’s short-term preferences. For long-term preferences, we introduce a target behavior-oriented gating module to guide the learning process, followed by a sequence capturer to uncover sequential patterns.
When constructing positive sample pairs, we treat the target behavior as a direct expression of short-term interests and apply the subsequence swapping strategy from \cite{MBASR} to generate new samples for long-term preferences. Finally, an adaptive preference-aware fusion module integrates both preferences to form the final user representation. Next, we will elaborate on these components.
% ===============

\section{METHODOLOGY}
\label{METHODOLOGY}
% In this section, we will introduce each component of our BDPL in detail.

\subsection{Embedding Initialization}
% 讲一下emb的初始化
Given a user's interaction sequence, we first transform the input one-hot encoding vectors of item $v_i$, behavior type $b_i$, and position $p_i$ into their corresponding embedding vectors $\boldsymbol{h}\in \mathbb{R}^{d\times 1}$, where $d$ denotes the embedding size. Specifically, the embeddings are initialized as:
% $x_{v_i}$, $x_{b_i}$ and $x_{p_i}$ are the one-hot vectors of the IDs of . 
\begin{equation}
    \boldsymbol{h}_{v_i}=x_{v_i} \boldsymbol{W}_v; \quad
    \boldsymbol{h}_{b_i}=x_{b_i} \boldsymbol{W}_b; \quad
    \boldsymbol{h}_{p_i}=x_{p_i} \boldsymbol{W}_p \quad
\end{equation}
where $\boldsymbol{W}_v \in \mathbb{R}^{|\mathcal{V}|\times d}$, $\boldsymbol{W}_b\in \mathbb{R}^{|\mathcal{B}| \times d}$, and $\boldsymbol{W}_p \in \mathbb{R}^{|\mathcal{S}_u| \times d}$ are learnable matrices. 

\subsection{Behavior-Aware Graph Construction}
% 
% Typically, target behaviors exhibit stronger user intent, which is the goal of our prediction. 
In an HSR scenario, different behaviors exhibit varying degrees of users' diverse interests, with target behaviors often showing stronger user intent. To better capture user preferences, we aim to design behavior-aware subgraphs to distinguish between auxiliary behaviors and target behaviors, with a greater focus on target behaviors, which is our prediction goal. Here, we construct heterogeneous graphs $\mathcal{G}={\{\mathcal G_e,\mathcal G_p }\}$ to mine complex multi-behavior patterns from the auxiliary-behavior perspective and the target-behavior perspective, respectively.

For each graph $\mathcal G_* = {\{\mathcal V,\mathcal{E_*} }\}$, $\mathcal{V}$ denotes the whole node set of items, and $\mathcal{E_*}$ represents the set of edges under the specific behavior, where $* \in \{e, p\}$.
Each triple ($v_i$, $r$, $v_j$) in $\mathcal{E_*}$ means that a user subsequently interacted with item $v_j$ after $v_i$, and $r$ denotes the behavior transition between these two consecutive items.
Since transitions within the same type of behavior provide valuable insights into user preferences, we first define the behavior-aware homogeneous relation to extract features from different behaviors, including from examination to examination $(e2e)$ and from purchase to purchase $(p2p)$. 

Additionally, we incorporate heterogeneous behavior transitions to capture multi-granularity behavior dynamics. Here, we introduce the behavior transition $e2p$ (from examination to purchase), as it effectively uncovers the transition relationships between the auxiliary behaviors and target behavior, aligning well with our prediction objectives. Unlike previous studies, we argue that transitions should extend beyond adjacent items, capturing multi-hop dependencies between multiple preceding auxiliary behaviors and the target behavior. This ensures that the model learns the full scope of user intent rather than focusing solely on the most recent $e2p$ transition, which may overlook critical long-range dependencies.

To determine a suitable range for the $e2p$ transition 
(i.e., the number of auxiliary behaviors preceding a purchase behavior), 
we segment an original sequence at purchase behaviors, creating some behavior-aware subsequences. Each subsequence starts with an auxiliary behavior and ends with a purchase, serving as the basic unit for capturing $e2p$ transitions.
For each target behavior $b_t$ in a user-item interaction sequence, we let $o$ be the index of the first auxiliary behavior in the current subsequence, and connect $v_t$ with all items $v \in \{v_{t'}\}_{o \leq t' < t}
$ under the $e2p$ transition. This enables us to fully capture the transitions from auxiliary behaviors to the target behavior within each subsequence.

Finally, let $\mathcal{R}_e$ and $\mathcal{R}_p$ represent the complete set of behavior transition types in $\mathcal{G}_e$ and $\mathcal{G}_p$, where $\mathcal{R}_e = \{e2e^+, e2e^-, e2p^-\}$ and $\mathcal{R}_p$ = $ \{p2p^+, p2p^-, e2p^+\}$. Here, ``+'' and ``-'' correspond to ``$\textit{forward}$'' and ``$\textit{backward}$''. Taking items $v_1$ and $v_2$ for example, $(v_1, e2e^-, v_2)$ denotes first examining $v_1$ and then $v_2$.
The synergy of these two aspects allows us to comprehensively extract valuable behavior patterns from both homogeneous and heterogeneous behavior transitions, providing deeper insights into user intentions. In the left half of Figure \ref{framework}, we also provide a detailed illustration of the different behavior transition relationships. 
Next, we will further describe how to encode these multiple relationships.

% \vspace{-0.2cm}
\subsection{Behavior-Aware Graph Encoding}
% 图卷积
% ===开头说明用了级联，从目标p到辅助e===
% 
For each item $v_i$ under the target behavior $p$, we aggregate the information from its neighboring items under a specific transition type $r$ to refine the item embedding by mean pooling:
% \vspace{0.1cm}
\begin{equation}
    \boldsymbol{h}^{(p, l)}_{v_i,r}=\frac{\sum_{v_j \in \mathcal{N}_r(v_i)} \boldsymbol{h}_{v_j}^{(p, l-1)}}{\left|\mathcal{N}_r(v_i)\right|}
\end{equation}
where $\mathcal{N}_r(v_i)$ denotes the neighbor set of item $v_i$ with the specific behavior transition type $r \in \mathcal{R}_p$, and $\boldsymbol{h}_{v_i}^{(p, 0)} = \boldsymbol{h}_{v_i} + \boldsymbol{h}_{b_i}$. Taking ${e2p}^+$ as an example, we can formalize the associated neighbor set as $\mathcal{N}_{e2p^+}(v_i)=\{v_j \mid\left(v_j, e2p+, v_i\right) \in \mathcal{E}_p \}$, where each triple denotes an arrow from item $v_i$ to item $j \in \mathcal{N}_r(v_i)$ in Figure \ref{framework}. Moreover,
to achieve a more fine-grained distinction between different types of behavior transitions, we apply an attention mechanism to adaptively assign weights to these transition relationships. Accordingly, we can generate the informative representation of item $v_i$ at layer $l$ under the target behavior $p$ via:
\begin{equation}
    % \setlength{\abovedisplayskip}{2pt}
    % \setlength{\belowdisplayskip}{2pt}
    % \small
    \boldsymbol{\alpha}_{v_i, r} =\frac{\exp \left(\boldsymbol{h}_{v_i,r}^{(p,l)} \boldsymbol{W}_ r^{(l)} / \sqrt{d}\right)}{\left.\sum_{r \in \mathcal{R}_p} \exp \left(\boldsymbol{h}_{v_i,r}^{(p,l)} \boldsymbol{W}_r^{(l)} / \sqrt{d}\right)\right)}
\end{equation}

\begin{equation}
    \boldsymbol{h}_{v_i}^{(p,l)}=\sum_{r \in \mathcal{R}_p} \boldsymbol{\alpha}_{v_i, r} \boldsymbol{h}_{v_i,r}^{(p,l)}
\end{equation}
where $\boldsymbol{W}_r^{(l)} \in \mathbb{R}^{d \times d}$ is the attention matrix. After $L$ layers' propagation, we aggregate the learned embeddings at each layer to form the final representation $\boldsymbol{h}_{v_i}^{(p)} \in \mathbb{R}^{d\times1}$, which can be formulated as:
% \vspace{-0.1cm}
\begin{equation}
    \boldsymbol{h}_{v_i}^{(p)}=\sum_{l=0}^L \boldsymbol{h}_{v_i}^{(p, l)}
\end{equation}

Since target behaviors indicate strong positive feedback signals, they are often considered highly significant in revealing user interests. These insights can assist in refining the understanding of auxiliary behaviors to better reflect user preferences. As illustrated in Figure \ref{framework}, the flow of data moves from the target behavior to auxiliary behaviors. After obtaining the representation of the target behavior, we feed it into the modeling of auxiliary behaviors. 
% By using the learned representations of items with target behaviors as the initial representations for items with auxiliary behaviors, we aim to enhance the representational power of auxiliary behaviors through the guidance provided by the target behaviors. 
Additionally, inspired by \cite{MB-CGCN}, to prevent the loss of behavior diversity that could result from directly using $\boldsymbol{h}_{v_i}^{(p)}$ as an input embedding for auxiliary behaviors, we perform a feature transformation on it as:
% 讲为什么要级联
% \vspace{-0.1cm}
\begin{equation}
    \boldsymbol{h}_{v_i}^{(e,0)}=\boldsymbol{W}_p \boldsymbol{h}_{v_i}^{(p)}
\end{equation}
As with the target behavior, we aggregate the preference information at the auxiliary behavior level in the same way, denoted as $\boldsymbol{h}_{v_i}^{(e)} \in \mathbb{R}^{d\times1}$. 

% Finally, we generate the multi-behavior embedding via:
% \begin{equation}
%     \boldsymbol{\tilde{h}}_{v_i} = \frac{1}{|\{e,p\}|} \sum_{*\in \{e,p\}}\boldsymbol{h}_{v_i}^{(*)}
% \end{equation}
Finally, we generate multi-behavior embeddings $\boldsymbol{\tilde{h}}_{v_i} \in \mathbb{R}^{d \times 1}$ by applying average pooling to the convolutional representations across all behaviors.

\subsection{Short-Term Preference Learning}

\subsubsection{Behavior-Integrated Attention Encoder}
With the behavior-aware graph encoder, we obtain a rich behavior-semantic representation for each item. We then stack all item embeddings from user $u$'s historical interaction sequence into a matrix $\boldsymbol{H} \in \mathbb{R}^{|\mathcal{S}_u| \times d}$. To further capture the dependencies between items as well as valuable sequential information, we feed $\boldsymbol{H}$ into stacked self-attention blocks (SAB), which consist of a \textit{multi-head self-attention} (MSA) layer and \textit{feed-forward networks} (FFN).
Specifically, $\boldsymbol{H}$ is first enriched with a learnable position matrix $\bm {P} = [\bm{h}_{p_1}, \bm{h}_{p_2}, \cdots, \bm{h}_{p_|\mathcal{S}_u|}] \in \mathbb{R}^{|\mathcal{S}_u| \times d}$ to model the influence of positions in the sequence, obtaining the final input matrix $\bm{X}^{(0)}$. We then pass $\bm{X}^{(k)}$ through an MSA($\cdot$) layer as follows: 
% \vspace{-0.08cm}
\begin{equation}
% \small
    % \setlength{\abovedisplayskip}{3pt}
    % \setlength{\belowdisplayskip}{3pt}
    {\rm MSA}(\boldsymbol{X}^{(k)}) = {\rm Concat(head_1,head_2,\cdots,head}_h) \boldsymbol{W}^O
\end{equation}
% \vspace{-2pt}
\begin{equation}
% \small
    % \setlength{\abovedisplayskip}{2pt}
    % \setlength{\belowdisplayskip}{2pt}
    {\rm head}_i = {\rm Att}(\bm{X}^{(k)} \bm{W}_i^Q, \bm{X}^{(k)} \bm{W}_i^K, \bm{X}^{(k)}\bm{W}_i^V)
\end{equation}
\begin{equation}
% \small
    \setlength{\abovedisplayskip}{1pt}
    \setlength{\belowdisplayskip}{1pt}
    {\rm Att}(\boldsymbol{Q, K, V}) = {\rm softmax}(\frac{\boldsymbol{QK}^T}{\sqrt{d/h}})\boldsymbol{V}
\end{equation}
where Att($\cdot$) is the scaled dot-product attention, using three linear transformation weight matrices $\bm{W}_i^Q, \bm{W_i^K}, \bm{W_i^V} \in \mathbb{R}^{d \times \frac{d}{h}}$, to map the input $\bm{X}^{(k)}$ into query, key and value vectors, respectively. $\bm{W}^O$ is the projection matrix of the output, $h$ is the number of heads, and $\sqrt{d/h}$ is the scale factor to avoid large values of the inner product. Then, we apply a position-wise feed-forward network with a ReLU activation function to enable the model to capture non-linear relationships. 
Moreover, we incorporate LayerNorm, dropout, and residual connection modules to mitigate the over-fitting problem:
\begin{equation}
% \small
    % \setlength{\abovedisplayskip}{2pt}
    % \setlength{\belowdisplayskip}{2pt}
    {\rm FFN}(\bm{X}^{(k)}_{Att}) = {\rm ReLU}(\bm{X}^{(k)}_{Att}\bm{W}_1 + \bm{b}_1)\bm{W}_2+\bm{b}_2
\end{equation}
\begin{equation}
% \small
    % \setlength{\abovedisplayskip}{2pt}
    % \setlength{\belowdisplayskip}{2pt}
    \bm{X}^{(k)} = {\rm LayerNorm}(\bm{X}^{(k)} + {\rm Dropout}({\rm FFN}(\bm{X}^{(k)}_{Att}))
\end{equation}

After $K_{SAB}$ attention blocks, we get the final output $\bm{X}\in \mathbb{R}^{|\mathcal{S}_u|\times d}$ and take the last row in $\bm{X}$ as the user $u$'s short-term preferences, denoted as $\bm{z}_u^s \in \mathbb{R}^{d\times1}$.

\subsubsection{Short-Term Contrastive Preference Learning}
When capturing users' short-term interests, most existing approaches typically rely on the most recently interacted item or the last $m$ items, which may overlook the shifts in user interests concerning the last item. Thus, we argue that directly using the user's next interaction item (i.e., the target item) to explicitly express short-term preferences can more accurately reflect the user's current interests. In this way, we can easily mine the user's short-term preferences from two different perspectives, i.e., the target item's representation $\bm{h}_{v_n}$ and the representation $\bm{z}_u^s$ learned by the encoder. We further employ contrastive learning to reinforce the modeling of short-term preferences. To be specific, we take $\bm{h}_{v_n}$ and $\bm{z}_u^s$ as a positive pair, and other target items in the same batch as negative samples, denoted as $\bm{H}_{v_n}^-$. The loss function $\mathcal{L}_{cl}^s$ can be defined as:
% \vspace{-0.1cm}
\begin{equation}
    % \small
    \footnotesize
    \mathcal{L}_{cl}^s (\bm{z}_u^s, \bm{h}_{v_n})= -{\rm{log}} \frac{{\rm{exp(sim}}(\bm{z}_u^s, \bm{h}_{v_n}))}{{\rm{exp(sim}}(\bm{z}_u^s, \bm{h}_{v_n})) + \sum\limits_{\bm{h}_{v_n}^-\in \bm{H}_{v_n}^-}{\rm{exp}}({\rm{sim}}(\bm{z}_u^s, \bm{h}_{v_n}^-))}
\end{equation}
where $n$ denotes the length of the historical interaction sequence, i.e., $|S_u| + 1$. sim($\cdot$) means the similarity function, for which we choose dot product to calculate the similarity between two representations. 
As a result, we can guide and enhance the learning of a user's short-term preferences in a target item-oriented manner.
% 短期偏好和target item 做对比学习
% 短期偏好，
\subsection{Long-Term Preference Learning}

% \subsubsection{Encoder}
\subsubsection{Target-Behavior-Oriented Gating}
Unlike short-term preferences, long-term preferences reflect a user's stable intentions from a more macro and holistic perspective. Here, we design different modules to explore a user's long-term preferences to avoid the interference caused by similar optimization objectives on preference learning from two perspectives. Specifically, to obtain more fine-grained and stable representations of a user's long-term preferences from the user’s perspective, we first design a target-behavior-oriented gating mechanism to 
fuse the effect of the target behavior on the user's historical behaviors,
\begin{equation}
   \bm{H'} = \bm{H} \otimes \sigma (\bm{H}\bm{W}_{g}^1 + \bm{W}_{g}^2\bm{h}_{b_{|\mathcal{S}_u|}}^T)
\end{equation}
where $\bm{W}_{g}^1 \in \mathbb{R}^{d \times 1}$ and $\bm{W}_{g}^2\in \mathbb{R}^{L\times d}$, $\sigma(\cdot)$ is the sigmoid function, $\otimes$ is the element-wise product, and $\bm{h}_{b_{|\mathcal{S}_u|}}$ is the target behavior embedding for user $u$.
\subsubsection{Long-Term Preference Capturer}
Considering that graph convolution cannot capture the sequential information, inspired by \cite{BMLP}, we adopt some stacked sequence capture blocks (SCB) to extract the sequence-level information. Specifically, the input dimension of SCB is the sequence length $|\mathcal{S}_u|$, and it captures the sequentiality between items by performing a nonlinear mapping on the sequence dimension. The output of the $k$-th SCB is as follows:
\begin{equation}
    \bm{H'}^{(k)}_{SCB} = \bm{H'}^{(k)} + ({\rm{GELU}}({\rm{LayerNorm}}(\bm{H'})^T\bm{W}_3)\bm{W}_{4})^T
\end{equation}
where $\bm{W}_3 \in \mathbb{R}^{|\mathcal{S}_u|\times d}$ and $\bm{W}_4 \in \mathbb{R}^{d \times |\mathcal{S}_u|}$ are learnable matrices. After $K_{SCB}$ sequence capture blocks, we are able to reveal the user's global preferences $\bm{H}_{SCB}\in \mathbb{R}^{|\mathcal{S}_u|\times d}$, which encompass both the behavior and sequential information. Finally, we obtain the user's long-term preferences $\bm{z}_u^l \in \mathbb{R}^{d\times 1}$ by averaging the representations of all items in the interaction sequence.

% \begin{equation}
%     \bm{h}_l = \frac{1}{|\mathcal{S}_u|}\sum_{i=1}^{|\mathcal{S}_u|}\bm{\tilde{h}}_{v_i}
% \end{equation}

\subsubsection{Long-Term Contrastive Preference Learning}
Long - term preferences are crucial for modeling user intent as they reflect a user's overall preferences over their entire lifecycle. In a prior work, MBASR \cite{MBASR} segments the original user interaction sequence $\mathcal{S}_u$ into multiple subsequences $\{s^1, s^2, \dots, s^m\}$ based on behavior types and introduces explicit data augmentation methods within and between these subsequences. 
Based on this, we adopt the subsequence swapping strategy to construct a positive sample sequence $\mathcal{\hat{S}}_u$ corresponding to the original sequence $\mathcal{S}_u$. Through contrastive learning, we aim to minimize the differences between augmented views of the same user's historical sequence and maximize the differences between sequences of different users.
Specifically, following MBASR, we first randomly select a $source$ subsequence and obtain its index $idx_{source}$. Then, we compute sampling probabilities for other subsequences based on their positional proximity to the \textit{source}, assigning higher weights to those closer to it. Then, a $destination$ subsequence is sampled based on this probability, and finally, we swap these two subsequences to obtain an augmented sequence $\mathcal{\hat{S}}_u$:
 \begin{equation}
        \mathcal{S}_u = \left\{s^1, \cdots, \underline{s}^{idx_{source}} , \cdots, \underline{s}^{idx_{dest}} , \cdots, s^m\right\} 
    \end{equation}
     % \vspace{-0.4cm}
    \begin{equation}
        \hat{\mathcal{{S}}}_u^{} = PS(\mathcal{S}_u) = \left\{s^1, \cdots, \underline{s}^{idx_{dest}} , \cdots, \underline{s}^{idx_{source}} , \cdots, s^m\right\} 
    \end{equation}

Then, we use the following loss function $\mathcal{L}_{cl}^l$ to distinguish the augmented representation of the same sequence from others:
% \vspace{-0.1cm}
\begin{equation}
\small
    \mathcal{L}_{cl}^l (\bm{z}_u^l, \bm{\hat{z}}_u^{l})= -{\rm{log}} \frac{{\rm{exp(sim}}(\bm{z}_u^l, \bm{\hat{z}}_u^{l}))}{{\rm{exp(sim}}(\bm{z}_u^l, \bm{\hat{z}}_u^{l})) + \sum\limits_{\bm{\hat{z}}_u^{l^-}\in \bm{\hat{Z}}_l^-}{\rm{exp}}({\rm{sim}}(\bm{z}_u^l, \bm{\hat{z}}_u^{l^-}))}
\end{equation}
where $\bm{z}_u^l$, $\bm{\hat{z}}_u^l \in \mathbb{R}^{d \times 1}$ are the long-term preferences of the original sequence and the augmented sequence, respectively. And the corresponding negative set is the long-term preferences of other sequences in the same batch, denoted as $\bm{\hat{Z}}_l^-$. 

\subsection{Prediction and Optimization}
\subsubsection{{Preference-aware Fusion}}
To encode different preference patterns from both long-term and short-term preferences, we design a preference gating module that can adaptively balance and integrate the preferences to derive the final interest:
\begin{equation}
% \setlength{\abovedisplayskip}{3pt}
% \setlength{\belowdisplayskip}{3pt}
% \begin{equation}
    \beta = \sigma([\bm{z}_u^s\Vert \bm{z}_u^l]\bm{W}_f)
\end{equation}
\begin{equation}
    \bm{o}_u = \beta\cdot\bm{z}_u^s + (1-\beta)\cdot\bm{z}_u^l
\end{equation}
where $\bm{W}_f \in \mathbb{R}^{2d\times 1}$ is the learnable fusing matrix, $\sigma$ denotes the sigmoid activation function, and $\Vert$ represents the concatenation operation.
\subsubsection{Prediction}
After obtaining the sequence representation $\bm{o}_u$, we first calculate the user $u$'s preference score towards each item $v_i \in \mathcal{V}$. A softmax function is then applied to normalize these scores, yielding the predicted probability that user $u$ will purchase item $v_i$ at the next time step:
% \vspace{-0.1cm}
\begin{equation}
% \small
% \setlength{\abovedisplayskip}{3pt}
% \setlength{\belowdisplayskip}{3pt}
    \hat{y}_i = {\rm{softmax}}(\bm{o}_u^T\bm{h}_{v_i})
\end{equation}
Then, we adopt the cross-entropy loss for parameter learning, which can be formulated as:
\begin{equation}
% \small
% \setlength{\abovedisplayskip}{3pt}
% \setlength{\belowdisplayskip}{3pt}
\mathcal{L}_{rec}=-\sum_{i=1}^{|\mathcal{V}|} \delta(v_i) \log \left(\hat{y}_i\right)
\end{equation}
where the indicator function $\delta(v_i)=1$ only if item $v_i$ is the true interacted item of user $u$ at the next time step, and $\delta(v_i)=0$ otherwise. 
% \subsubsection{Dual-Channel Contrastive Objective}
Accordingly, we adopt multi-task learning to optimize the ranking loss and the dual-channel contrastive preference learning loss, and the final joint loss is a linear weighted sum:
\begin{equation}
    \mathcal{L} = \mathcal{L}_{rec} + \lambda_1 \mathcal{L}_{cl}^s + \lambda_2 \mathcal{L}_{cl}^l 
\end{equation}
where $\lambda_1$ and $\lambda_2$ are the hyper-parameters to control the weight of two contrastive learning tasks.
\subsubsection{Time Complexity Analysis}
The time complexity of SASRec and our BDPL is O($|\mathcal{S}_u|d^{2} + |\mathcal{S}_u|^{2}d$), which demonstrates the efficiency of our BDPL. Some specific analysis is detailed as follows: Our BDPL mainly consists of three modules. For the behavior-aware graph construction and encoding module, the time complexity is O($(|\mathcal{R}_{e}| + |\mathcal{R}_{p}|)|\mathcal{S}_u|d^{2}$) = O($(6|\mathcal{S}_u|d^{2}$). For the short-term preference learning module, the time complexity is O($|\mathcal{S}_u|d^{2} + |\mathcal{S}_u|^{2}d$). As for the long-term preference learning module, the time complexity is O($|\mathcal{S}_u|d^{2}$). In all, the time complexity of our BDPL is O($8|\mathcal{S}_u|d^{2} + |\mathcal{S}_u|^{2}d$) = O($|\mathcal{S}_u|d^{2} + |\mathcal{S}_u|^{2}d$), which is the same as that of SASRec.
\section{EXPERIMENTS}
In this section, we first present our experimental setup in detail. Then, we conduct extensive experiments to answer the following key research questions (RQs): \textbf{Q1:} How does our BDPL perform compared with various recommendation methods?
\textbf{Q2:} How does each designed component contribute to the performance of our BDPL?
\textbf{Q3:} How effective is the cascade network structure driven by the target behavior in our BDPL?
\textbf{Q4:} How do different hyper-parameter settings impact the performance of our BDPL?

\subsection{Experimental Settings}
\subsubsection{Datasets} We conduct extensive experiments on three datasets widely used in the recommender systems community to evaluate the effectiveness of our BDPL, i.e., \textbf{Tmall\footnote{https://tianchi.aliyun.com/dataset/dataDetail?dataId = 42}}, \textbf{UB\footnote{https://tianchi.aliyun.com/dataset/dataDetail?dataId = 649}} and \textbf{JD\cite{BVAE}}. These three datasets are collected from different e-commerce platforms and contain rich heterogeneous user behaviors.
Identical to previous studies \cite{BAR}, we preprocess these datasets as follows: i) we discard the cold-start items with fewer than 10 interaction records in UB, and 20 in Tmall and JD; ii) we discard the cold-start users with fewer than 10 interaction records in Tmall, and 5 in UB and JD; iii) for duplicated (user, item, behavior) triple in a sequence, we only keep the first one; and iv) we adopt the widely used leave-one-out strategy in sequential recommendation to split each dataset into three parts, i.e., the last purchase interaction as the test data, the penultimate purchase interaction as the validation data, and the rest as the training data. The statistics of the processed datasets are shown in Table \ref{tab:dataset}.
% \vspace{-0.3cm}
\begin{table}[!htbp]
\centering
\setlength{\abovecaptionskip}{0pt}
    \caption{Statistics of the processed datasets.}
    % \vspace{-0.4cm}
    \label{tab:dataset}
    \scalebox{0.8}
    {
    \begin{tabular}{lccccc}
        \toprule
        % \midrule
        \textbf{Dataset} & \textbf{\#Users} &\textbf{\#Items} & \textbf{\#Examinations} & \textbf{\#Purchases} & \textbf{Avg. length}\\
        \midrule
        Tmall & 17,209 & 16,177 & 446,442 & 223,265 & 62.29\\
        UB & 20,858 & 30,793 & 470,731 & 136,250 & 29.10\\
        JD & 11,367 & 12,266 & 131,298 & 75,774 & 16.16\\
        \bottomrule
    \end{tabular}}
\end{table}
\begin{table*}[!htbp]
    \renewcommand\arraystretch{1}
    \caption{Performance comparison of all methods on the three datasets in terms of HR and NDCG. The best results are boldfaced, and the second-best results are underlined.}
    \vspace{-0.3cm}
    \label{RQ1}
    \centering
    \scalebox{0.7}{
    \begin{tabular}{c|l|cccc|cccccccc|ccc|c}
    % \hline
        \toprule
        \multicolumn{1}{c}{\multirow{1}{*}{Dataset}} & \multicolumn{1}{c}{\multirow{1}{*}{Metric}} & \multicolumn{1}{c}{{GRU4Rec}} & \multicolumn{1}{c}{{Caser}} & \multicolumn{1}{c}{{NextItNet}} & \multicolumn{1}{c}{{SASRec}} & \multicolumn{1}{c}{{RIB}} & \multicolumn{1}{c}{{BAR}} & \multicolumn{1}{c}{{GPG4HSR}} & \multicolumn{1}{c}{{NextIP}} & \multicolumn{1}{c}{{BMLP}} & \multicolumn{1}{c}{{GHTID}} & \multicolumn{1}{c}{{M-GPT}} & \multicolumn{1}{c}{{GEAR}} & \multicolumn{1}{c}{{CL4Rec}} & \multicolumn{1}{c}{{DuoRec}}  & \multicolumn{1}{c}{{DCRec}} & \textbf{Ours} \\ 
        \midrule
        \multirow{6}{*}{Tmall} & HR@5 & 0.0530  & 0.0250 &0.0537 & 0.0643  & 0.0566  & 0.0706  & 0.0676  & 0.0748  & 0.0740 & 0.0354 & 0.0479 & 0.0436 & 0.0700  & 0.0726  & \underline{0.0761}  & \textbf{0.0818}  \\ 
        ~ & HR@10 & 0.0710  & 0.0370 & 0.0705& 0.0871  & 0.0772  & 0.0927  & 0.0905  & 0.0984  & 0.0983 & 0.0542 & 0.0686 & 0.0646 & 0.0927  & 0.0963  & \underline{0.0984}  & \textbf{0.1052}  \\ 
        ~ & HR@20 & 0.0905  & 0.0553   &0.0908 & 0.1101  & 0.0991   & 0.1144  & 0.1163  & 0.1230  & \underline{0.1243} & 0.0730 & 0.0923 & 0.0904 & 0.1180  & 0.1189  & 0.1208  & \textbf{0.1294}  \\ 
        ~ & NDCG@5 & 0.0355  & 0.0161  &0.0372 & 0.0435 & 0.0380  & 0.0451 & 0.0518 & 0.0501  & 0.0490 & 0.0230 & 0.0314 & 0.0278 & 0.0490  & 0.0515  & \underline{0.0538}  & \textbf{0.0562}  \\ 
        ~ & NDCG@10 & 0.0413  & 0.0199   &0.0426 & 0.0509 & 0.0447  & 0.0562 & 0.0525 & 0.0577  & 0.0595 & 0.0284 & 0.0381 & 0.0346 & 0.0563  & 0.0592  & \underline{0.0611}  & \textbf{0.0634}  \\ 
        ~ & NDCG@20 & 0.0462  & 0.0246  &0.0477 & 0.0567 & 0.0502 & 0.0616 & 0.0590 & 0.0639  & 0.0652 & 0.0337 & 0.0440 & 0.0411 & 0.0627  & 0.0649  & \underline{0.0667}  & \textbf{0.0699}  \\ 
        \midrule
        \multirow{6}{*}{UB} & HR@5 & 0.0314  & 0.0355  &0.0295 & 0.0534  & 0.0447  & 0.0526  & 0.0596  & 0.0583  & 0.0624 & 0.0256 & 0.0280 & 0.0402 & 0.0655  & \underline{0.0707}  & 0.0690  & \textbf{0.0718}  \\ 
        ~ & HR@10 & 0.0472  & 0.0544 & 0.0445 & 0.0762  & 0.0652 & 0.0784  & 0.0820  & 0.0863  & 0.0861 & 0.0419 & 0.0436 & 0.0599 & 0.0885  & \underline{0.0952}  & 0.0921  & \textbf{0.0990}  \\ 
        ~ & HR@20 & 0.0677  & 0.0792  &0.0637 & 0.1020  & 0.0906   & 0.1059  & 0.1107 & 0.1168  & 0.1233 & 0.0628 & 0.0656 & 0.0855 & 0.1162  & \underline{0.1248}  & 0.1217  & \textbf{0.1288}  \\ 
        ~ & NDCG@5 & 0.0208  & 0.0224  & 0.0192 & 0.0353 & 0.0297   & 0.0341 & 0.0391 & 0.0378  & 0.0444 & 0.0160 & 0.0176 & 0.0255 & 0.0459  & \textbf{0.0507}  & \underline{0.0500}  & {0.0499}  \\ 
        ~ & NDCG@10 & 0.0259  & 0.0285  &0.0241 & 0.0426 & 0.0364  & 0.0424 & 0.0464 & 0.0469  & 0.0521 & 0.0213 & 0.0227 & 0.0319 & 0.0534  & \underline{0.0576}  & 0.0575  & \textbf{0.0588}  \\ 
        ~ & NDCG@20 & 0.0311  & 0.0347 & 0.0289 & 0.0491 & 0.0427  & 0.0494 & 0.0537 & 0.0545  & 0.0616 & 0.0266 & 0.0281 & 0.0384 & 0.0604  & \underline{0.0651}  & 0.0649  & \textbf{0.0663}  \\ \midrule
        \multirow{6}{*}{JD} & HR@5 & 0.2200 & 0.2252  &0.2166 & 0.2409 & 0.2332  & 0.2235  & 0.2418  & 0.2382  & 0.2029 & 0.1956 & 0.2201 & 0.2015 & {0.2471}  & 0.2435  & \underline{0.2476}  & \textbf{0.2509}  \\ 
        ~ & HR@10 & 0.2979  & 0.3032  & 0.0907 & 0.3151  & 0.3196  & 0.3068  & 0.3233  & 0.3323  & 0.3102 & 0.2646 & 0.3045 & 0.2941 & 0.3483  & 0.3456  & \underline{0.3522}  & \textbf{0.3557}  \\ 
        ~ & HR@20 & 0.3773  & 0.3834  & 0.3660 & 0.3951 & 0.4013  & 0.3833  & 0.3980 & 0.4149  & 0.4275 & 0.3318 & 0.3858 & 0.3857 & 0.4350  & 0.4377  & \underline{0.4417}  & \textbf{0.4428}  \\ 
        ~ & NDCG@5 & 0.1420 & 0.1453   &0.1367& 0.1434 & 0.1452 & 0.1364 & {0.1449} & \underline{0.1455}  & 0.1306 & 0.1272 & 0.1405 & 0.1235 & 0.1418  & 0.1428  & 0.1442  & \textbf{0.1459}  \\ 
        ~ & NDCG@10 & 0.1673  & 0.1736  & 0.1607 & 0.1698 & 0.1712   & 0.1634 & 0.1789 &\underline{0.1790}  & 0.1655 & 0.1495 & 0.1678 & 0.1535 & 0.1756  & 0.1763  & 0.1782  & \textbf{0.1799}  \\ 
          & NDCG@20  & 0.1874   & 0.1938  & 0.1791 & 0.1901  & 0.1999  & 0.1827  & 0.2002  & 0.2000   & 0.1925  & 0.1665 & 0.1884 & 0.1768 & 0.1975   & 0.1996   & \underline{0.2009}   & \textbf{0.2020}   \\ 
          \bottomrule
    \end{tabular}}
\end{table*}
% \vspace{-1cm}
\subsubsection{Evaluation Metrics}
We employ two popular metrics to evaluate the performance of all methods, i.e., $\textit{Hit Ratio}$ (HR@N) and $\textit{Normalized Discounted Cumulative Gain}$ (NDCG@N), where $N \in \{5, 10, 20\}$. In this paper, we report the ranking results over the entire item set (i.e., full ranking) to avoid the bias introduced by sampling \cite{full}.
% \vspace{-0.3cm}
\subsubsection{Baseline Methods.}
To demonstrate the effectiveness of our BDPL, we compare it with fifteen representative baselines that can be classified into three groups. 
% {\color{blue}These baselines are briefly introduced in related work.}

\textbf{Homogeneous Sequential Recommendation:}
    (1) \textbf{GRU4Rec} \cite{GRU4Rec}: An RNN-based model that first applies GRU to model user sequences for session-based recommendation.
    (2) \textbf{Caser} \cite{Caser}: A CNN-based model that adopts horizontal and vertical convolutional operations to model high-order patterns.
    (3) \textbf{NextItNet} \cite{NextItNet}: A CNN-based model that utilizes stacked CNNs to increase the receptive field to model long-range item sequences effectively.
    (4) \textbf{SASRec} \cite{SASRec}: An attention-based model that uses the multi-head attention mechanism to learn sequence representations.

\noindent\textbf{Heterogeneous Sequential Recommendation:}
    (1) \textbf{RIB} \cite{RIB}: It combines the item embedding and behavior embedding as the input of a GRU layer, followed by an attention layer to capture the influence of different behaviors on user preferences.
    (2) \textbf{BAR} \cite{BAR}: It designs a behavior-aware attention layer to model the relationship between the target and current behaviors.
    % , and leverages the next time step's behavior for more accurate recommendation.
    (3) \textbf{GPG4HSR} \cite{GPG4HSR}: It constructs a global graph to capture behavior transitions and personalized graphs to enrich the sequence representation with contextual information.
    (4) \textbf{NextIP} \cite{NextIP}: It applies a target-behavior aware context aggregator to mine knowledge specific to different behaviors.
    (5) \textbf{BMLP} \cite{BMLP}: It is an MLP-based model that uses one module to capture multi-granularity behavior type relationships and another module to aggregate auxiliary behavior subsequences.
    (6) \textbf{GHTID} \cite{GHTID}: It proposes an item-item co-occurrence graph and an item-item transition graph to capture global and local heterogeneous behavior transitions, respectively. 
    (7) \textbf{M-GPT} \cite{M-GPT}: It constructs an interaction-level graph and a novel multi-faceted Transformer to model interaction-level dependencies and extract multi-grained preferences, respectively. 
    (8) \textbf{GEAR} \cite{GEAR}: It captures temporal dynamics by introducing a time-bias term. 
    
\noindent\textbf{Contrastive Sequential Recommendation}:
    (1) \textbf{CL4SRec} \cite{CL4SRec}: It utilizes three augmentation operators to generate contrastive samples, including item crop, mask and reorder.
    (2) \textbf{DuoRec} \cite{DuoRec}: It proposes both the model-level augmentation and the supervised sampling strategy to learn better sequence representations.
    (3) \textbf{DCRec} \cite{DCRec}: It leverages an adaptive conformity-aware augmentation strategy to reduce the impact of popularity bias.
% \vspace{-0.5cm}
\subsubsection{Implementation Details}
We implement our BDPL with PyTorch and use Adam for optimization. For fair comparison, the hyper-parameters of both the baseline methods and our BDPL are well-tuned with grid search on the validation set. To be specific, we set the embedding size to 64, the batch size to 128, the dropout rate to 0.5, and fix the sequence length to 50 for all methods. 
For multi-head self-attention-based methods (i.e., SASRec, BAR, NextIP, CL4SRec, DuoRec, DCRec, M-GPT and GEAR), the number of heads and layers are both tuned in $\{1,2,3\}$. For GNN-based methods (i.e., GPG4HSR, GHTID and DCRec), the number of graph convolution layers is chosen from $\{1,2,3\}$.For Caser, we set the number of horizontal and vertical filters as 16 and 4, respectively, and choose the height of the horizontal filters from $\{2, 3, 4\}$. For CL4SRec and DuoRec, the weight of the CL loss is chosen from $\{0.1, 0.2, 0.3, 0.4, 0.5\}$. For DCRec, the mean value of the conformity scores is searched from $\{0.3, 0.4, 0.5, 0.6, 0.7\}$, the weight for the self-supervised learning loss and the Kullback-Leibler divergence loss are searched from $\{$5e-4, 1e-3, 5e-3, 1e-2$\}$ and $\{$1e-3, 1e-2, 1e-1, 1$\}$, respectively. For our BDPL, we select the GNN depth from $\{1, 2, 3\}$. Both the numbers of SAB and SCB are tuned in the range of $\{1, 2, 3\}$. The weight $\lambda_1$ for the short-term CL loss and $\lambda_2$ for the long-term CL loss are searched from $\{$1e-3, 1e-2, 1e-1, 5e-1$\}$.
In the training process, we train the model with a learning rate of 1e-3 and get the best model via early stopping w.r.t. HR@10 on the validation set. 
% For all baselines, other hyper-parameters are searched according to the suggestions of the corresponding papers. For our BDPL, we select the GNN depth from $\{1, 2, 3\}$. Both the numbers of SAB and SCB are tuned in the range of $\{1, 2, 3\}$. The weight $\lambda_1$ for the short-term CL loss and $\lambda_2$ for the long-term CL loss are searched from $\{$1e-3, 1e-2, 1e-1, 5e-1$\}$.
% In the training process, we train the model with a learning rate of 1e-3 and get the best model via early stopping w.r.t. HR@10 on the validation set. 
% \vspace{-0.3cm}

\subsection{Overall Performance Comparison (RQ1)}
We report a comprehensive performance comparison between the baselines and our BDPL across different datasets in Table \ref{RQ1}. From the results, we can summarize the following main observations:
\begin{itemize}[leftmargin=*]
    \item Our BDPL consistently yields superior performance on all three datasets and almost all metrics, which clearly demonstrates its effectiveness and superiority in modeling multi-behavior heterogeneity and sequential dynamics. Specifically, the average improvements over the strongest baselines range from 0.25\% to 17.61\% on HR and NDCG.We attribute these gains to the behavior-aware graph encoder and the dual-channel contrastive preference learning paradigm: (1) The former effectively captures cross-type behavior dependencies and leverages target behaviors to guide the prediction of the next item. (2) The latter enhances the learning of both long-term and short-term user intentions through carefully designed positive pairs.
    \item For the homogeneous sequential recommendation methods, SASRec beats all the other three RNN- or CNN-based models on almost all datasets, which demonstrates the superior ability of the self-attention mechanism in modeling sequential patterns. Furthermore, SASRec outperforms even the MBSR models RIB on Tmall and UB, further proving its effectiveness in capturing item dependencies.
    \item For the heterogeneous sequential recommendation methods, the overall performance is better than that of homogeneous sequential recommendation models (except SASRec), which reveals the importance of modeling behavior types and their transition relationships. 
    It is worth noting that GHTID does not perform well across all three datasets. We believe this is because, during our data processing, auxiliary behaviors between the last two purchases are not retained. GHTID relies on the auxiliary behaviors preceding purchases to construct the global graph, which thus led to a sharp decline in performance. Similar experimental results have also been observed in M-GPT and GEAR.
    \item For the contrastive learning (CL)-based sequential recommendation methods, DCRec outperforms DuoRec on both Tmall and JD, suggesting that popularity bias does have a negative impact when augmenting representation learning by CL.
    \item Our BDPL shows a more significant improvement over the second-best model on Tmall compared with that on the other two datasets, indicating its greater advantage in capturing long-range item dependencies.
\end{itemize}

\subsection{Ablation Study (RQ2)}
\begin{figure}[!htbp]
    \centering 
    \includegraphics[width=0.5\textwidth]{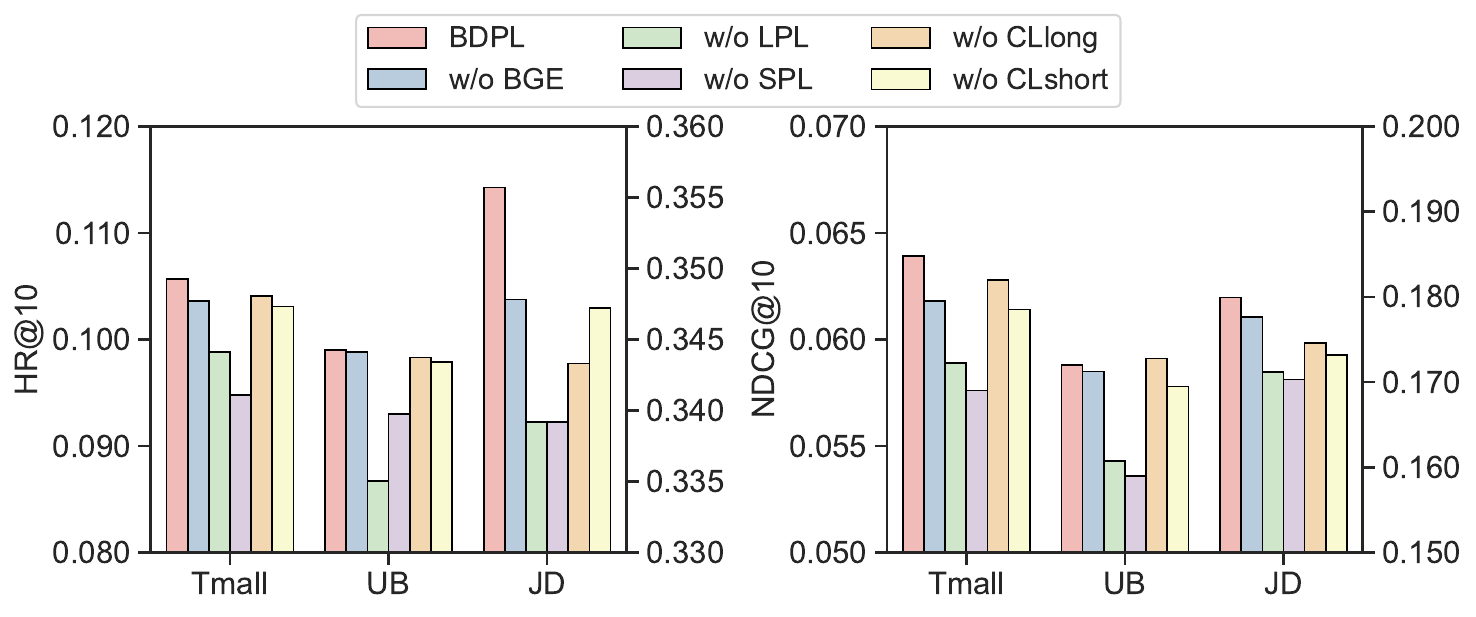}
    % \vspace{-0.05cm}
    \caption{Ablation study on key components of our BDPL. The left y-axis is shared by Tmall and UB, and the right y-axis is for JD.} 
    \label{RQ2}
\end{figure}

To verify the effectiveness of each designed component in our BDPL, we conduct an ablation study of the model with its five variants on Tmall, UB and JD, including: 1) {\bf{\textit{w/o}} \bf{BGE:}} we remove the behavior-aware graph encoder and directly use the initial embedding of items for preference learning. 2) {\bf{\textit{w/o}} \bf{SPL:}} we extract only a user's long-term preferences as the final representation. 3) {\bf{\textit{w/o}} \bf{LPL:}} we rely solely on a user's short-term preference information for the final representation. 4) {\bf{\textit{w/o}} $\bf{CL}_{short:}$} we disable the contrastive enhancement mode for short-term preferences. 5) {\bf{\textit{w/o}} $\bf{CL}_{long:}$} we do not perform contrastive learning on the long-term preferences. As shown in Figure \ref{RQ2}, due to the significant difference in metric scales between JD and the other two datasets, we adopt two separate vertical axes for better visual clarity and aesthetics. Specifically, the left y-axis is shared by Tmall and UB, while the right y-axis is dedicated to JD. 
From the results, we can obtain the following observations:
% \vspace{-0.5cm}

\begin{itemize}[leftmargin=*]
    \item Firstly, it is clear that the performance of our BDPL declines noticeably when any component is removed in most cases. This strongly indicates that each key component we designed contributes significantly to the overall performance improvement.
    \item Secondly, when only long-term or short-term preference learning is retained, the model experiences the greatest performance drop, which aligns with our design motivation. By combining both long-term and short-term preferences, we gain a deeper understanding of user intent, allowing us to view user behavior more comprehensively from a global and dynamic perspective.
    \item Thirdly, the effectiveness of the behavior-aware graph encoder is also intuitive. It learns item representations for various behavior types hierarchically and employs a cascade approach, allowing the target behavior to refine the modeling of auxiliary behaviors, which is crucial for capturing dependencies between heterogeneous behaviors.
    \item Finally, we observe that dual-channel contrastive learning effectively enhances preference representation learning, and their combination also produces complementary effects. This demonstrates that the two sets of explicit positive sample pairs we designed are both reasonable and effective.
    % 针对短期偏好和长期偏好的对比学习都取得了效果，并且将这两者结合能够达到互补的效果
\end{itemize}
\subsection{Study of Cascaded Flow Direction (RQ3)}
\label{Cascaded}
To demonstrate the effectiveness and rationale of the cascade network structure we designed, which focuses on the direction from the target behavior to auxiliary behaviors in the behavior-aware graph encoder (BGE), we conduct a more detailed analysis through experiments. Specifically, in our BGE, we employ cascaded network structures from the target behavior to auxiliary behaviors (denoted as $\bm{tar2aux}$) and from auxiliary behaviors to target behavior (denoted as $\bm{aux2tar}$) to examine the performance on HR@10 and NDCG@10 across three datasets. The results are shown in Table \ref{e2p}. It is evident that the recommendation performance w.r.t. $\bm{tar2aux}$ surpasses those w.r.t. $\bm{aux2tar}$ across nearly all metrics on all the three datasets, which aligns with our expectations. Compared with the auxiliary behaviors, the target behaviors more directly reflect users' true preferences and intentions, providing purer and more indicative user preference information. In contrast, auxiliary behaviors are often mixed with noisy and less reliable data. Therefore, we propose that using the target behavior graph convolution output as input for auxiliary behavior learning can better guide the model while mitigating noise effects.
% \vspace{-0.5cm}
\begin{table}[htbp]
  \centering
  \caption{Performance comparison of our BDPL with different cascaded methods, where the best results are highlighted in bold.}
    \vspace{-0.2cm}
\scalebox{0.75}{
    \begin{tabular}{clcccccc}
    \toprule
    \multirow{2}{*}{Method} & \multicolumn{2}{c}{Tmall} & \multicolumn{2}{c}{UB} & \multicolumn{2}{c}{JD} \\
    \cmidrule{2-7} & HR@10 & NDCG@10 & HR@10 & NDCG@10 & HR@10 & NDCG@10 \\
    \midrule
    $aux2tar$ & 0.1046 & 0.0634 & 0.0977 & \textbf{0.0590} & 0.3471 & 0.1709 \\
    $tar2aux$ & \textbf{0.1052} & \textbf{0.0635} & \textbf{0.0990} & 0.0588 & \textbf{0.3557} & \textbf{0.1799} \\
    \bottomrule
    \end{tabular}}
   \label{e2p}
\end{table}
% \vspace{-0.5cm}
\subsection{Parameter Sensitivity Study (RQ4)}
\begin{figure}[htbp]
    \centering
    \setlength{\abovecaptionskip}{0.05cm}
    \begin{subfigure}[b]{\columnwidth}
        \centering
        \includegraphics[width=0.9\linewidth]{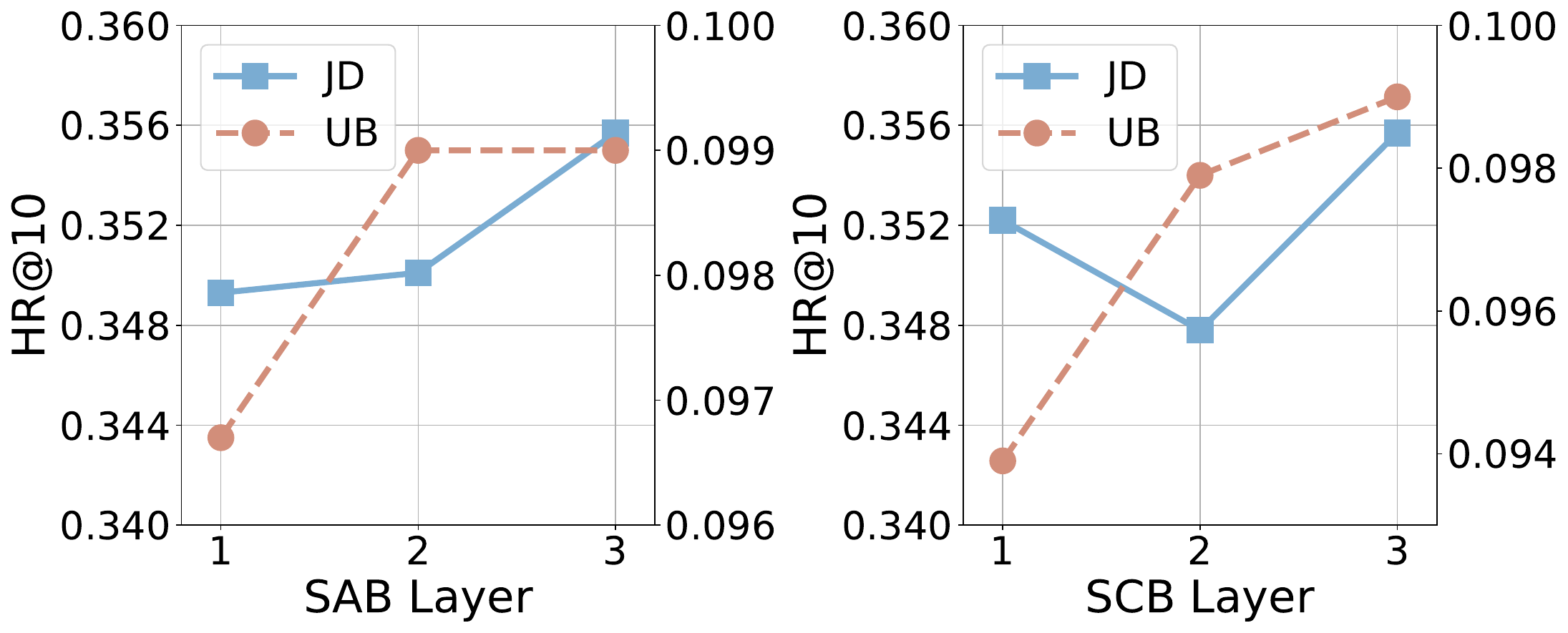}
        % \caption{Sensitivity on parameter 1.}
        % \label{fig:RQ3_1}
    \end{subfigure}
    \vspace{0.1cm}
    \begin{subfigure}[b]{\columnwidth}
        \centering
        \includegraphics[width=0.45\linewidth]{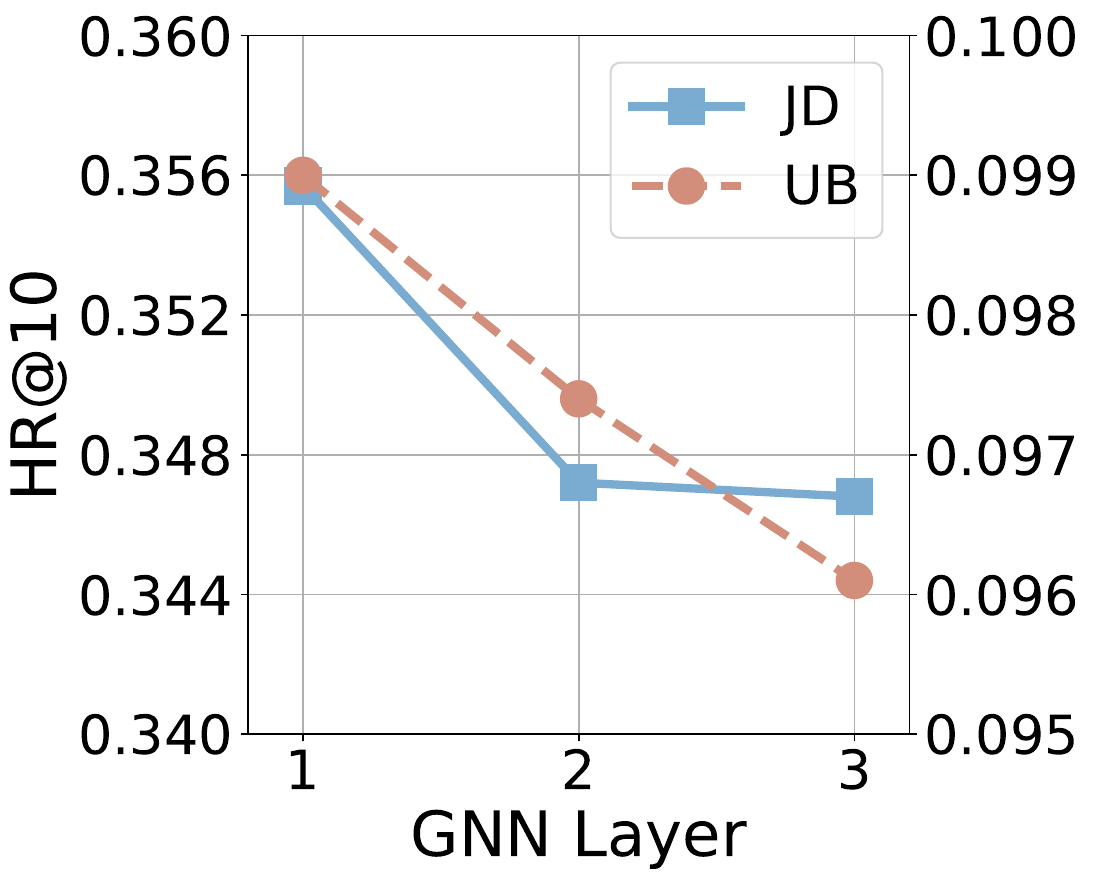}
        % \caption{Sensitivity on parameter 2.}
        % \label{fig:RQ3_2}
    \end{subfigure}
    \caption{Study of parameter sensitivity of our BDPL.}
    \label{RQ3}
\end{figure}

We further perform a parameter sensitivity analysis on JD and UB to show the impact of the numbers of graph propagation layers $L$, self-attention blocks $K_{SAB}$ and sequence capture blocks $K_{SCB}$. Empirically, we conduct a search for the optimal values of these three key hyper-parameters from the set $\{1, 2, 3\}$. The results are reported in Figure \ref{RQ3}. Similar to the setup in RQ2, two separate vertical axes are employed to better visualize the metric differences across datasets. We can find that:

As the numbers of self-attention blocks and sequence capture blocks increase, the HR@10 metric on both datasets continues to improve. This is due to the ability of additional layers to learn more complex feature representations and strengthen information transfer, thus capturing the underlying patterns in the data more effectively. More SAB helps deepen the model's understanding of contextual relationships, while SCB mines sequence information, which is critical for modeling long-time dependencies. Conversely, when the number of GNN layers is set to 1, the model performs satisfactory. This phenomenon can be attributed to the rich input features of the designed behavior-aware graph structure, which allow a single-layer GNN to aggregate information while mitigating the risk of overfitting effectively. 
\section{CONCLUSIONS AND FUTURE WORK}
In this paper, we present a novel behavior-aware dual-channel preference learning framework (BDPL) for heterogeneous sequential recommendation. Firstly, we construct behavior-aware subgraphs to capture the diverse behavior transition patterns in user-item interactions, followed by a behavior-aware graph encoder to learn user interests oriented towards the target behavior. Secondly, we dynamically fuse and extract a user's core interests from the perspectives of long-term and short-term preferences, and develop two explicit contrastive learning tasks to enhance the two types of representations. Finally, we adaptively merge the long-term and short-term preferences to obtain the final preference representation for the user. We believe that the preference-level contrastive learning paradigm empowers the model with a stronger ability to exploit the contextual information of user preferences, resulting in superior performance.
Empirical results on three real-world datasets validate the strengths of our BDPL when competing with the state-of-the-art recommendation methods.

For future works, we are interested in extending our BDPL for multi-modality heterogeneous sequential recommendation by leveraging LLMs.

\bibliographystyle{ACM-Reference-Format}
\bibliography{ref}

% \appendix









\end{document}